\documentclass[review]{elsarticle}

\usepackage{lineno,hyperref}
\usepackage{booktabs}
\usepackage{multirow}
\usepackage[table]{xcolor}
\usepackage{lipsum}
\usepackage{changepage}
\usepackage{amsmath}
\usepackage[noend]{algpseudocode}
\usepackage[noend]{algpseudocode}
\usepackage{subfig}
\usepackage{rotating}
\usepackage[linesnumbered,ruled,vlined]{algorithm2e}

\modulolinenumbers[5]
\newcommand{\ophs}{\ensuremath{\mathrm{OP_\mathrm{HS}}}\xspace}
\newcommand{\hsprms}{\ensuremath{\mathrm{HS\mbox{-}PRMS}}\xspace}
\newcommand{\pahs}{\ensuremath{\mathrm{PA_{HS}}}\xspace}
\newcommand{\hsp}{\ensuremath{\mathrm{hs_{prms}}}}
\newcommand{\dirug}{\ensuremath{\mathrm{directUG}}\xspace}
\newcommand{\effhp}{\ensuremath{\mathrm{effectiveHS_{prms}}}\xspace}
\newcommand{\userSub}{\ensuremath{\mathrm{userSub}}}
\newcommand{\s}{\ensuremath{\mathrm{S}}}
\newcommand{\U}{\ensuremath{\mathrm{U}}}

\journal{
Future Generation Computer Systems (FGCS)} 

\begin{document}

\begin{frontmatter}

\title{Next-Generation Big Data Federation Access Control: A Reference Model}

\author[mymainaddress]{Feras M. Awaysheh\corref{mycorrespondingauthor}}
\cortext[mycorrespondingauthor]{Corresponding author: feras.awaysheh@usc.es}
\author[mythirdaddress]{Mamoun Alazab}
\author[mysecondaryaddress]{Maanak Gupta}
\author[mymainaddress]{Tom{\'a}s F. Pena}
\author[mymainaddress]{Jos{\'e} C. Cabaleiro}

\address[mymainaddress]{Centro Singular de Investigaci\'on en Tecnolox\'ias Intelixentes (CiTIUS), University of Santiago de Compostela, Spain}
\address[mythirdaddress]{College of Engineering, IT \& Environment, Charles Darwin University, Australia}
\address[mysecondaryaddress]{Department of Computer Science, Tennessee Tech University, Cookeville, TN, USA}

\begin{abstract}

This paper discusses one of the most significant challenges of next-generation big data (BD) federation platforms, namely, Hadoop access control. Privacy and security on a federation scale remain significant concerns among practitioners. Hadoop's current primitive access control presents security concerns and limitations, such as the complexity of deployment and the consumption of resources. However, this major concern has not been a subject of intensive study in the literature. This paper critically reviews and investigates these security limitations and provides a framework called BD federation access broker to address 8 main security limitations. This paper proposes the federated access control reference model (FACRM) to formalize the design of secure BD solutions within the Apache Hadoop stack. Furthermore, this paper discusses the implementation of the access broker and its usefulness for security breach detection and digital forensics investigations. The efficiency of the proposed access broker has not sustainably affected the performance overhead. The experimental results show only 1\% of each 100 MB read/write operation in a WebHDFS. Overall, the findings of the paper pave the way for a wide range of revolutionary and state-of-the-art enhancements and future trends within Hadoop stack security and privacy.

\end{abstract}


\end{frontmatter}

\nolinenumbers

\section{Introduction}
\label{intro}

Apache Hadoop~\cite{Hadoop}, the BD landmark, has become a large-scale data analytics operating system. The large community behind Hadoop has been working to improve its stack to meet the increasing demands and requirements of BD. 
Enterprises across all major industries have adopted Hadoop due to its capability to store and process an abundance of new types of data and leverage modern data architecture. With a broad spectrum of both structured and unstructured workloads, Hadoop abstracts the computing resource management, task scheduling, and data management, while maintaining a satisfactory level of security and isolation.

The Hadoop distributed file system (HDFS)~\cite{borthakur2007hadoop} is typically deployed as part of a large-scale Hadoop platform to support commodity hardware and accommodate different processing frameworks. It is utilized to handle data management and access to the Hadoop ecosystem using a master/slave architecture. It is also successfully employed by several distributed systems and can be used by different resource schedulers as a data storage system, e.g., HTCondor~\cite{HTCondor} and Spark~\cite{zaharia2010spark}. IAM in the HDFS ecosystem can be defined as the set of tools and mechanisms that enables end users and applications to interact securely with system core functionalities, thus ensuring appropriate access to data across the cluster. This security discipline can be separated into three abstraction layers: identification and access control, authentication, and authorization.

As more services and users have joined the Hadoop federation portfolio in pursuit of a scalable BD hub, access control has become increasingly critical. One of the main obstacles in the development of an adaptive access control solution for BD platforms is the lack of a standard model to which access control rules and the associated enforcement monitor can be bound. A recent study~\cite{colombo2018access} indicates that BD, as an emerging research trend, lacks standardized models for unifying user/application access to resources, services, and data. Moreover, Hadoop’s primitive configuration lacks any classification mechanism that improves metadata governance or facilitates auditing procedures. Therefore, formalizing the IAM features of Hadoop 3.x, in addition to emphasizing the complexity of securely utilizing their core functionality, is becoming a research interest. This trend serves as a form of knowledge capture by mapping current technologies of concern and formulating the latest Hadoop capabilities related to access control frameworks and audit log management.

This paper highlights the need for robust access control that handles the authentication and authorization operations within a federation scale. The paper defines the associated requirements of secure BD runtime and management of the Hadoop HDFS federation. It therefore proposes a reference model (RM) that provides a basis for building an interoperable data federation scheme that includes all of the major stages and reflects specifics in access control within Hadoop clusters using modern open-source technologies. Furthermore, this paper explains how the proposed models can be implemented using modern BD infrastructure (BDI) to meet the increasing demand for scalable access controls and digital forensics (using access log analysis). The key question we will be asking to address the IAM challenge is as follows: “How can we create a dynamic reference model that is compatible with the foreseen BDI scenarios in a federated setting?” We address this question by employing the proposed RM in a novel access broker that provides a single sign-on environment. The proposed BD Federation (BDF) broker is an access control logic component to securely connect the external users with the Hadoop cluster gateway. The work presented in this paper may be employed in a Hadoop federation environment across multi-tenant BD clouds, as well as within on-premise data centres.

An essential application of the proposed solution deals with security auditing and analysis of audit logs of BD operations. These log files contain detailed information about all access call activities regarding BD processes and infrastructure to be protected by centralizing the access control of the data lake, BD services, and underlying resources in a unified access broker pattern. This broker abstracts the identification, authentication, and authorization discussion within large-scale data analytics. This layer may also afford granular insights into pieces of information by performing security and risk assessment, tracking data pipeline audit logs, and examining behavioural analytics to meet their compliance and governance demands within the Hadoop 3.x platform.

The rest of this paper is organized as follows. Section \ref{overview} provides an overview of the Hadoop federation and its role in supporting BD operations. It also highlights complementary security measurements for BD frameworks and discusses the related work. The HDFS access control primitives and federation access provisioning limitations on which this study is based are investigated in Section \ref{AcessControl}. Section \ref{RM} describes the logic language for defining the RM, while we implement the proposed model in a novel access broker proof-of-concept framework in Section \ref{Implementation}. Section \ref{Audit} demonstrates the advantages of the proposed centralized audit log management, and we ultimately provide the conclusion in Section \ref{Summary}. 


\section{Hadoop Federation}
\label{overview}

As the size of a Hadoop cluster increases, the pressure on the NameNode and the ResourceManager increases. To alleviate this problem, Hadoop has introduced both HDFS Federation~\cite{HDFSFeb} and YARN Federation~\cite{YARNFeb}. HDFS Federation consists of the usage of multiple independent NameNodes, where each NameNode is responsible for managing a subset of the whole namespace. To understand how this federated architecture works, we need to explain how the single HDFS is designed. HDFS is composed of two main layers:

\begin{itemize}
    \item The namespace layer, which runs in the NameNode and is in charge of storing all information related to directories, files, and blocks (creation, deletion, modification, listing).
    \item The block storage service, which supports low level block-related operations and contains two parts:
    \begin{itemize}
        \item Block management, which runs in the NameNode and is responsible for block creation, deletion, modification, replication, and location.
        \item Physical storage, which runs in the DataNodes and is responsible for storing the data blocks and providing read/write access to them.
    \end{itemize}
\end{itemize}
    \begin{figure}[!t]
        \centering
        \includegraphics[width=0.7\textwidth]{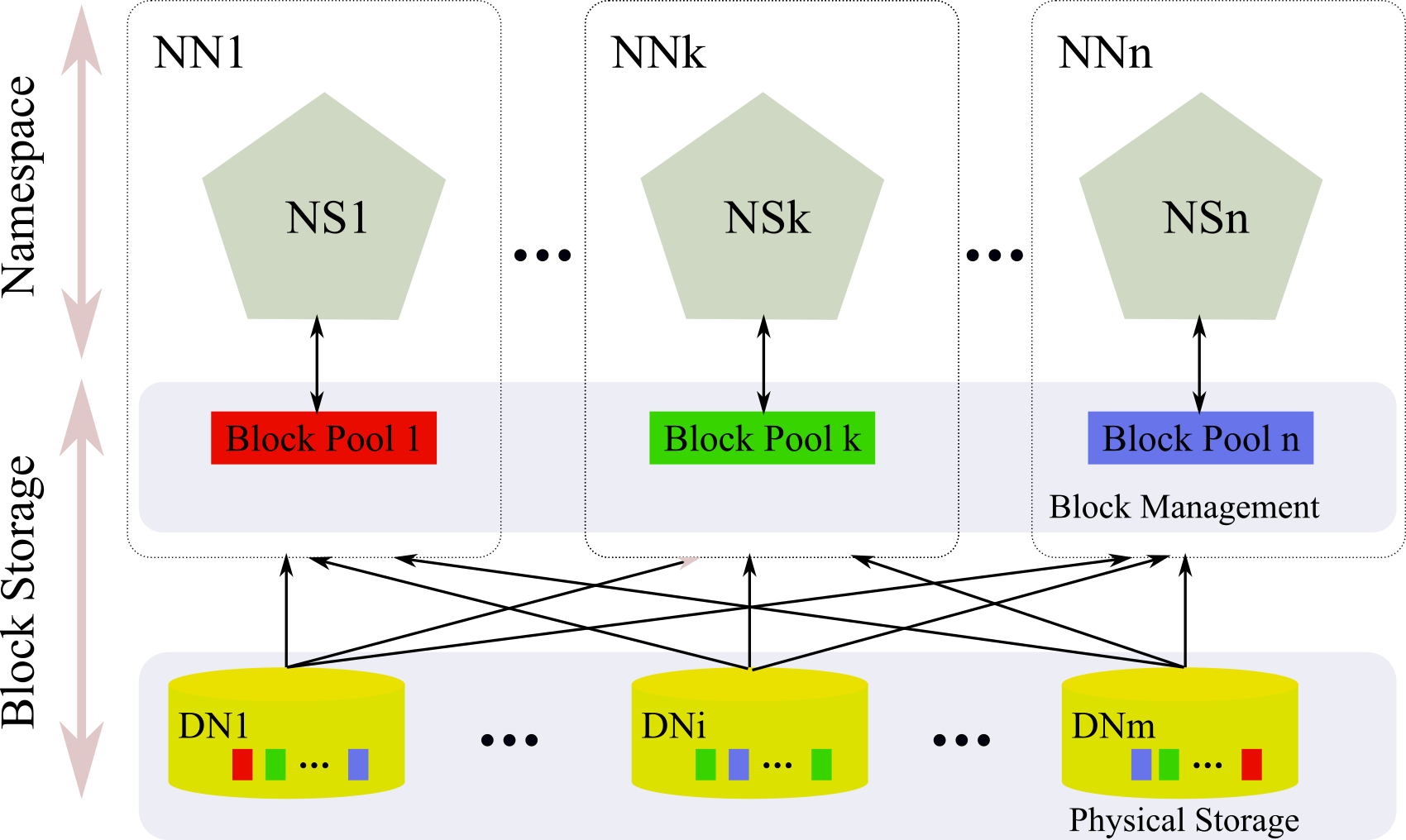}
        \caption{HDFS Federation Architecture with n NameNodes (NN) and Namespaces (NS) and m DataNodes (DN)}
        \label{fig:1}
    \end{figure}

Whereas physical storage can be easily scaled horizontally, simply by adding more DataNodes, previous versions of HDFS only allowed the namespace layer to be scaled vertically, as it runs on a single NameNode. The HDFS federation removes that limitation, allowing the namespace layer to be scaled horizontally. It uses a federation of multiple NameNodes that are independent; i.e., they do not require coordination with each other. Each NameNode is assigned a set of blocks, which is called a block pool. Different blocks in a Block Pool can live in any DataNode, and so each DataNode must register with all NameNodes in the cluster, sending them periodic heartbeats and reports about block status. A schema of this architecture is shown in Figure~\ref{fig:1}.


However, on Jan 16, 2019, Apache Hadoop released its 3.2.0 stable platform, while the first stable 3.x line was released on Apr 6, 2018. This release incorporates several significant enhancements over the previous primary release line. A federated architecture has also been proposed for Apache YARN. This approach allows YARN to scale to tens of thousands of nodes. In this architecture, a large YARN cluster is split into a set of sub-clusters. Each sub-cluster has its own ResourceManager and compute nodes. From the point of view of applications, the federated cluster is still seen as a single huge YARN cluster, and tasks can be scheduled on any node of the cluster. The federation system is responsible for negotiating with the resource managers of the sub-clusters and providing resources to each application. YARN Federation functionality relies on reliable connectivity across sub-clusters to deploy atop the HDFS Federation. A federation architecture across data centres among several physical confines requires further investigation.

\subsection{HDFS federation}
\label{HDFSfederation}

In this section, we present a brief overview of HDFS federation architecture and highlight the Apache Hadoop security features.

\subsubsection{HDFS architecture}
\label{architect}

HDFS~\cite{shvachko2010hadoop} is a distributed and open-source file system designed to meet the rapidly growing demands of large-scale data management and access. The HDFS is composed of two primary daemons: (i) a single NameNode (NN) that is deployed at the cluster master node and (ii) several DataNodes (DNs) running at the cluster slaves (usually one per node). The NN runs the namespace process, which manages the file system information and regulates access to files using a conventional hierarchical organization. In addition, it tracks block locations and numbers and writes log information files for auditing purposes. Internally, a file is partitioned into multiple data blocks that are placed into various DNs to be stored in local disks (block storage). These blocks are replicated for fault tolerance over several DNs from different racks (if possible). The NN makes all decisions regarding replicas and periodically receives a static heartbeat; the default period is every 3 seconds. The NN checks the expiry time report of a heartbeat every 200 seconds (as a default timeout). When a new file is updated, the NN places replicas of the file blocks in different nodes and racks (if available). 

Conventionally, the Hadoop cluster runs several DNs, but it has only one NN (and one namespace) for all DNs (HDFS High-Availability allows running two NNs when working in active/standby mode). The DNs can be scaled both vertically (by adding more resources to the nodes) or horizontally (by adding more nodes to the cluster). The NN, however, may only scale vertically, which means that one needs to add more resources (CPUs and RAM) to that NN server to serve more DN metadata. Notice that metadata are preserved in memory for minimizing latency and enabling faster retrieval. This approach causes a single point of failure and, hence, limits the number of blocks, files, and directories maintained on the file system. The HDFS Federation has been introduced to cope with this issue. 

To enable a universal block storage layer, Hadoop performed separation of namespace and blocked storage~\cite{HDFSFederation}. A federation BD environment, through multi-independent namespaces for block management and a common block pool for data storage, improves scalability and isolation of Hadoop operations. Therefore, by loosening the tightly coupled block storage and namespace, each DN registers with all the NNs in the cluster (this increases the authentication requirement). These DNs send periodic heartbeats, block reports and handle commands from the NNs. This allows horizontal scaling of the NN and enables the aggregation of geo-distributed Hadoop clusters. This feature directly enhances throughput by adding more access enforcers (typically, NNs in HDFS architecture), which improves read/write operations. 

Moreover, the HDFS federation services high-intensity BD applications that block vast resources on the NN by distributing them among different namespaces. However, this imposes authentication and authorization challenges, as well as security concerns, which we address in the next section. For instance, a federation cluster may improve the query performance in a Hive framework -- as Apache Hive manages data in partitions within different tables and locations. This setting can store various tables in separate namespaces, or even save the table partitions in different namespaces. In principle, this optimizes load balancing across multiple namespaces (e.g., one for archival data and another for current data), which reduces each namespace load and improves the application isolation.

\subsection{Complementary security measures for HDFS federation}
\label{security}

\begin{figure}[!t]
  \centering
  \includegraphics[width=.7\linewidth]{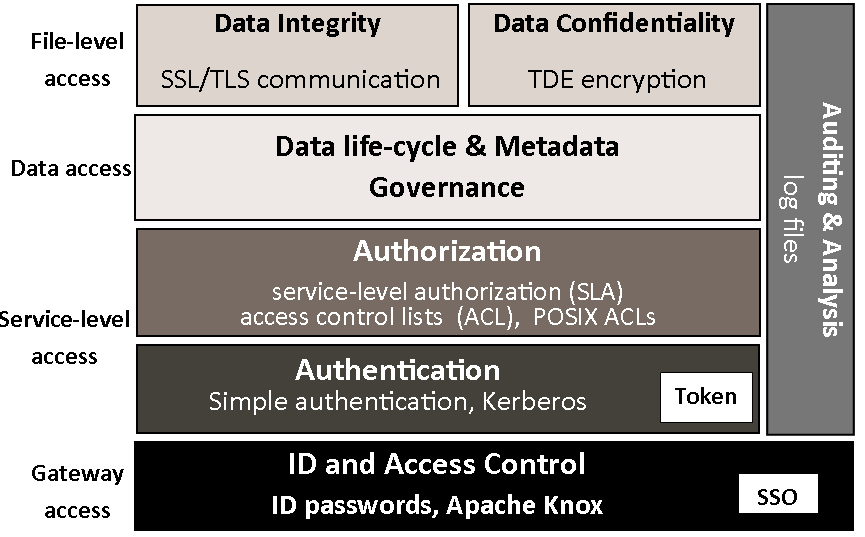}
  \caption{Apache Hadoop security layer abstraction.}
  \label{fig:2}
\end{figure}

Figure~\ref{fig:2} presents a security layered abstraction of the engineering BD system, which involves several essential functions. The first three layers, identification (ID) and access control, authentication, and authorization, are combined, representing the IAM processes of the Apache Hadoop cluster and the objective of this paper. The other security layers (data governance, integrity, confidentiality, and security auditing) are briefly presented in this section to perform a comprehensive security discussion of a BD environment.    

\begin{itemize}
\item {\bf Access control} is a service gateway to securely and efficiently communicate with the BD federation. This layer verifies the external client's access to the system using the user identification (ID) and passwords. Every client username and IP address must be in the host file or in a DNS table, and must match the client-given password. This process may also include Apache Knox~\cite{ApacheKnox}, a unified gateway framework for Hadoop services and ecosystem that can be utilized as a single-sign-on (SSO) gateway.

\item {\bf Authentication} is the act of confirming authentication access to the Hadoop services and HDFS data (after user log-on to the cluster), i.e., the process of actually determining the client identity. In a non-secure Hadoop mode, this layer is disabled and internal entities, i.e., clients (confirmed by the host OS), application, and ecosystem, interact directly with the Hadoop services. However, Hadoop secure mode~\cite{SecureMode} enables an authentication mechanism to verify that an entity is what it claims to be using the Kerberos protocol (authentication based on tokens). Each Hadoop service and user must be authenticated by the Kerberos keytab file (binary containing the information needed to log) using Hadoop tokens to initialize trust between a client/application and the HDFS (more details in Section~\ref{limitation}). Authentication for access to the Hadoop services web console requires enabling the HTTP SPNEGO protocol as a backend for Kerberos credentials~\cite{Spnego}. Thus, preventing the stored data in HDFS from unauthorized access is applied both to all clients accessing the cluster and to any service claimed to be part of the cluster. 

\item {\bf Authorization} is the process of defining access rights which an entity (service, daemon, or client) can perform to the given service. It manages access in the context of a specific service, resource, and data functionality provided by the cluster. Hadoop service level authorization (SLA) manages the fundamental set of permissions, such as defining the users and groups who are authorized to make service calls (e.g., data access) to that service. The call will pass the authorization check only if the user making the call belongs to an authorized service entity. It also provides fine-grain access control (table, column, and file levels) by enforcing the access control list (ACL), i.e., consistent policy administration across all Hadoop ecosystems~\cite{SLA}. ACL combines three elements: effect (allow or deny), action (e.g., data access or execution), and resources (e.g., NameNode1, Hive table). In principle, authorization is considered to be the final IAM layer in a Hadoop security abstraction, which means that no additional security mechanism is required as an IAM intention for an authorized client. Nevertheless, data encryption both at rest and in transition is still required, in addition to security analysis, auditing, and metadata governance, which represent high-level security services when combined as in Figure~\ref{fig:2}.      

\item {\bf Data Governance} is a capability that ensures adequate manageability of data through the complete lifecycle (i.e., data at rest, movement, and processing). This layer includes several dimensions, including classification (labeling and description), source tracking, and quality across data sources. Providing a typical store for exchanging metadata tags and attributes among the Hadoop stack can be achieved using Apache Atlas~\cite{atlas} and Apache Solr~\cite{solr} for defining data types and fields using full-text indexing and querying techniques.

\item {\bf Data Integrity and Confidentiality} is a capability to ensure adequate consistency and accuracy of data-at-rest as well as in-transit. This security layer includes validity and recoverability approaches, as Hadoop 3.x utilizes erasure coding for fault tolerance ~\cite{Erasure}. However, aiming for confidentiality, HDFS implements end-to-end encryption with so-called transparent data encryption~\cite{TransparentDataEncryption}. This HDFS encryption occurs at the file level of on-disk data and is stored as NN metadata. On the other hand, Hadoop wire-security (such as for data transfer between Web-console and client) is managed via SSL/TLS for HTTP communications.  

\item {\bf Security auditing and analysis:} The aggregation of log files and reports provides a robust audit capability within different components of the Hadoop ecosystem. This layer may also afford granular insights into pieces of information by performing security and risk assessment, tracking data pipeline audit logs, and examining behavioural analytics to meet their compliance demands within Hadoop.

\end{itemize}

\subsection{Related work}
\label{related}

The security of BD deployment architectures and Hadoop service trust have always been labelled as research concerns. Several research papers have discussed the Hadoop ecosystem privacy and access management~\cite{gupta2017multi,gupta2017object,gupta2018attribute,colombo2017enhancing, report2,report1,das2011adding,o2009hadoop,sharma2014securing}. The access control requirements and privacy analysis of Hadoop frameworks have been addressed in the literature~\cite{colombo2018access,colombo2015privacy}. Recently, Gupta et al.~\cite{gupta2017multi} and~\cite{gupta2017object} formalized the access control features of Hadoop core capabilities, as well as other security associated frameworks including Apache Ranger and Sentry. They extend their work with the HeABAC~\cite{gupta2018attribute} model, a multi-layer attribute-based access control model~\cite{gupta2016mathrm,jin2012unified,gupta2019dynamic} that provides fine-grained authorization policies for Hadoop. Ulusoy et al. proposed an approach for fine-grained access control authorization for MapReduce systems in~\cite{ulusoy2014vigiles,ulusoy2015guardmr}. Big data privacy issues are also well addressed, and novel solutions have been proposed in \cite{colombo2015privacy,lu2014toward,soria2016big,tene2012big}, in addition to an SSO framework for Hadoop services in~\cite{awaysheh2019poster}. Colombo et al. conducted a comprehensive study of big data technologies, including access control requirements, state-of-the-art and future trends, in \cite{colombo2019access,colombo2018access,colombo2017enhancing}.   
One of the earliest works by Kulkarni \cite{kulkarni2013fine} targets wide-column NoSQL databases which support content and context-based access control policies at different levels of the data model, such as row or column. Another work on Cassandra datastore \cite{shalabi2017cryptographically} presents a cryptographic enforcement of RBAC \cite{sandhu1996role} policies. 


Access control of data and resources from multiple sources within centralized computing systems has been a subject of intensive studies in the literature. Context-awareness using fuzzy logic conditions as access control approach has been proposed in ~\cite{kayes2019context} to address the dynamic outsourcing environment on the edge of the network, and for the intelligent transportation systems ~\cite{awaysheh2019big}. Aiming to protect the redundant data stored over the cloud, an approach by Zhou, Yukun et al. ~\cite{zhou2018similarity} supported flexible access control with revocation. Adopting a proxy re-encryption policy to update the process to the outsourced cloud was reported for BD deployments ~\cite{fugkeaw2018scalable} to support ciphertext re-encryption. Controlling access to sensitive BD, e.g., healthcare records, using the quantum mechanical equivalent of digital signature was introduced ~\cite{qiu2018quantum}, along with an unaddressed challenge of the data transfer process. Other than the implementations mentioned above, the BDs of time series (e.g., time-series databases) are required to be associated with an encrypted timestamp, the authors in ~\cite{noury2019access} proposed an access and inference control model to enforce time and value-based constraints over the hierarchical time series data. Meanwhile, the analyzing of malicious codes and intrusion detection for cybersecurity and malware detection was reported \cite{alazab2015profiling,alazab2019deep, vinayakumar2019deep,huda2016hybrids}  

Standardizing the security development schema of secure systems was first reported in~\cite{basin2006model}. This standardization unifies practices and languages for modelling security and access control among different implementers. A reference model for developing cloud applications was reported~\cite{hamdaqa2011reference}, along with a survey that supports federated access control~\cite{jie2011review}. Researchers from the National Institute of Standards and Technology (NIST) \cite{hu2014access} have presented a general access control model for BD processing frameworks which introduces the novel notion of chain of trust among entities to authorize access requests. Barker, S. proposed meta-model and best-practices of access control modelling~\cite{barker2009next}. Both the work of Barker and this research aim at unifying access control models by drafting a reference model. However, herein, we address the modern Hadoop 3.x status given the broad diversity of existent access control uniformity associated with a federation environment. Additionally, this paper distinguishes itself by addressing the HDFS federation access control challenges and maps cutting-edge frameworks to that problem domain.


\section{Access control in HDFS}
\label{AcessControl}
A client needs to perform tasks through the NN as a central policy enforcer for the HDFS. The NN receives the client call and allows it to reach data files which are stored in local disks via the DN pool. HDFS applications need a write-once-read-many access model for files. Once a file is created and written, it cannot be modified without an authorization access level policy. A secure IAM model must ensure that entities (clients, processes, and daemons) are who they claim to be (authentication) and that they have permissions to perform the requested operation (authorization).

\subsection{HDFS replica selection}

It is important to direct the reader's attention to ``how it works'' before presenting the access challenges within the federation settings.

\begin{table}[!t]
\centering
\caption{HDFS data access priority scheme.}
\label{table1}
\begin{center}
\def\arraystretch{1}%
\begin{tabular}{ccccc}
\toprule  
\rowcolor{lightgray} \textbf{Cluster} & \textbf{Rack} &  \textbf{Node} &  \textbf{Path} & \textbf{Priority} \\
\rowcolor{lightgray} \textbf{Number} & \textbf{Number} &  \textbf{Number} &   &  \\
\midrule
  \multirow{3}*{Cluster 1} & \multirow{2}*{Rack 1} & Node 1 & Local & First  \\
\cline{3-5}
& & Node 2 & Node 2\textgreater Rack 1 & Second \\
\cline{2-5}
       & Rack 2 & Node 3 & Node 3\textgreater Rack 2\textgreater Cluster 1 & Third \\
\bottomrule
\end{tabular}
\end{center}
\end{table}

The HDFS cluster responds to client calls via the closest block replica to the reader node, aiming to keep data movement and read latency to a minimum. Thus, the preferred replica is the one stored locally (if available) to the reader node or the replicas of the same rack to reduce bandwidth consumption (see Table~\ref{table1}). If the HDFS spans multiple rack clusters, then a replica placed in the local rack is preferred over any remote replica~\cite{shafer2010hadoop}. The NN manages this fault tolerance operation by implementing a rack-aware policy (e.g., the first replica is placed on the local node, the second is located in a different rack, and the third one is stored on another node within the local rack). By analysing the DN length to the parent, the NN selects the replica placement path. Consider the example in Table~\ref{table1} that assists in the further discussion concerning HDFS access control and HDFS Federation (section 5). A Hadoop application will use blocks locally within its current node. However, if the block has to be copied from a remote DN, it will calculate and pass the block to the clients in the order of proximity.

\begin{figure}
    \centering
    \begin{minipage}{0.5\textwidth}
        \centering
        \includegraphics[width=1.04\textwidth]{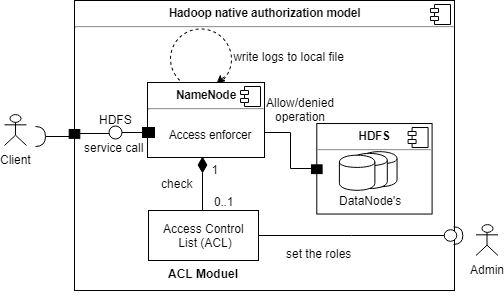} 
        \caption{Authorization access pattern in a \newline native Hadoop Yarn cluster.}
        \label{fig: 3}
    \end{minipage}\hfill
    \begin{minipage}{0.5\textwidth}
        \centering
        \includegraphics[width=.9\textwidth]{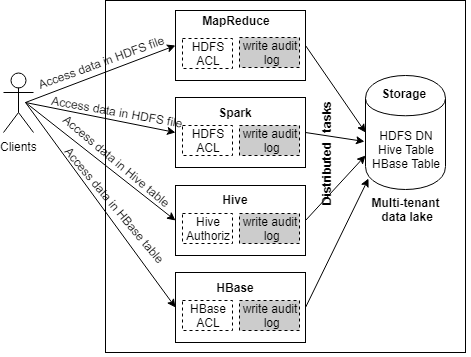} 
        \caption{Access pattern in a multi execution-engine Hadoop cluster with no central policy service.}
        \label{fig: 4}
    \end{minipage}
\end{figure}

Figure~\ref{fig: 3} demonstrates the HDFS authorization model pattern with an HDFS service call scenario. At first, Apache Yarn client make an HDFS call through the NN cluster as the service level access control. The NN checks the ACL, which had been set by the administrator, and the service call is either granted authorization to proceed with the operation or to deny it. Finally, the NN writes audit log information to a local file. System administrators must set this security mechanism for each Hadoop component and execution engine (Spark, Hive, HBase, etc.). Every ACL has one NN (and only one), while the NN could or could not have an ACL. In the last case, any service call made by the user that reaches the NN will gain access to DNs (the user still needs the ID and password for the external log).

\subsection{Apache Hadoop federation access management}
\label{AccessManage}

A cluster with high-level data analysis, i.e., sequel operations, and different execution engines will contact its security module (local policy authority or access enforcer) to validate access control rules. Figure~\ref{fig: 4} illustrates a Hadoop Yarn cluster with several data access requests managed by different BD execution frameworks. The cluster administrators need to separately create the ACLs by specifying authorization access based on the local policy authority (e.g., HDFS ACL, Hive ACL, etc.) for each framework~\cite{SLA}. The NN will perform a permission check before issuing any operation; those ACLs are located and defined in the \texttt{\$HADOOP\_CONF\_DIR/hadoop-policy.xml} file. In this case, the request will reach the local authorization module that allows/denies the operation and writes its own logs to a local file. These individual access enforcers (see Figure~\ref{fig: 4}), coupled to file permissions and different Hadoop framework orchestration, make it challenging and time-consuming to set a secure BD environment, not to mention the new BD federation features and multi-tenant BDaaS cloud deployment architecture. A solution relies on a cross-service authorization framework that provides a centralized policy authority to store and manage security policies for multiple ecosystem components.


\subsection{Limitations of BD Federation access control}
\label{limitation}

The broad participation nature of both clients and execution frameworks that are associated with the Hadoop federation warns against security concerns. In addition to the complexity of settings, dynamically and elastically scaling client authorization and policy provisioning are challenging tasks. A central policy enforcer to manage the access discussions seems mandatory. This section outlines the inherited limitations of implementing HDFS access control in a federation setting. 

Restricting access to the data stored in HDFS requires ensuring an accurate level of access by both clients and applications to the ecosystem components. In this way, by enabling Hadoop secure mode, every client, service, daemon, and process (which we refer to as entities) running within Hadoop must authenticate its membership of the cluster. The Hadoop underlying authentication service is based on Kerberos encapsulation, a leading local area network certificate-based protocol. Kerberos is a three-way protocol that requires every task to be authenticated before accessing the data. Therefore, only verified HDFS DN could register with the NN, and every task must be authenticated via the three-way handshake process before resource acquisition. This approach can cause performance degeneration in distributed clusters (e.g., think of a MapReduce application with thousands of tasks).

One solution to this problem is to complement Kerberos with Delegation Tokens (DT), a lightweight authentication process. With DT, a client or application is first authenticated with Kerberos, and DT is then applied for subsequent and iterative service calls. Thus, the cluster services and daemons are only authenticated with Kerberos, while the client's subsequent service calls and job tasks will use the DT credentials to interact with the HDFS. 
The use of Kerberos and DT in secure HDFS federation authentication is described in Figure~\ref{fig:5}:

\begin{enumerate}
\item Clients (and every entity) are initially confirmed to contact the NN using Kerberos authentication, and each client receives a DT certificate from the NN server to access the HDFS DNs using block tokens (access tickets) for every block with which they must communicate.

\item The client uses the access ticket for subsequent service calls, instead of using Kerberos. Furthermore, the distributed tasks use that DT certificate on behalf of clients to securely reach the NN at runtime.

\item The client and the distributed tasks can access the HDFS DN using the granted access tickets, which declare that the caller has the stated access rights to the data blocks within the job submission.
\end{enumerate}

A single-sign on (SSO) architecture can be managed using Kerberos with the lightweight directory access protocol (LDAP), which is enabled by setting \texttt{Hadoop.security.group.mapping.ldap.url} to true. Alternatively, implementing a single gateway that handles client access management behind a firewall can be achieved by employing Apache Knox. Knox will allow/deny users to access the ecosystem services before interacting with the Hadoop cluster. Following user verification using the Knox gateway, Knox will use its Kerberos principals to securely confirm access with other Hadoop services and daemons. 

\subsection{Problem definition}
\label{Problem}

\begin{figure}[t]
  \centering
  \includegraphics[width=.65\linewidth]{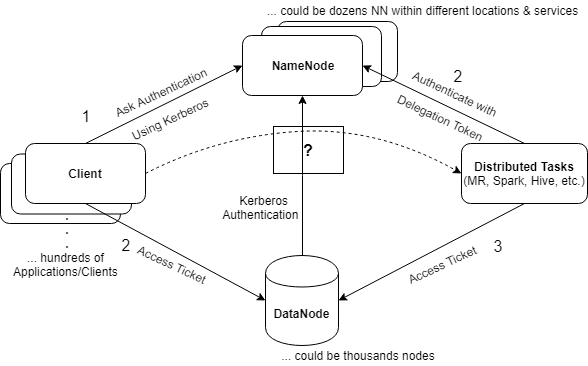}
  \caption{A simplified diagram of the HDFS federation authentication mechanisms.}
  \label{fig:5}
\end{figure}

The previous example demonstrates a single client contacting the NN, while several authentication mechanisms must be sequentially acquired for each Hadoop entity (daemon, client, and task). The DT ticketing process considerably decreases the amount of authentication requests; however, we argue that a federation Hadoop cluster can process the login requests of thousands of entities within a short period. A federation operation may require authorization checks for different components of the path: not only the final entity (as for a multi-tier federation architecture) 
This serialization of authentication before issuing the task will degenerate the system throughput in a large-scale data cluster and promptly become a bottleneck. This bottleneck may involuntarily cause a distributed denial of service attack within a federation environment. In addition, information logs are stored and managed locally (by NNs) without any classification mechanism that improves metadata governance or facilitation of auditing procedures.    

Relying on Hadoop IAM as demonstrated in figures~\ref{fig: 4} and~\ref{fig:5} could present the following underlying limitations:  

\begin{enumerate}
\item Access control complexity: Hadoop core IAM requires a client to go through multi-permission stages for each service that it is willing to contact. This results in additional time and bandwidth consumption in a federation setting. Additionally, the fact that Kerberos/DT tickets are time-limited means that we need to re-authenticate the entities based on the given maximum lifetime. A client must request a new DT and pass it to the running process. 

\item Long-running services: Due to token expiration time, supporting long-running jobs beyond the token maximum lifetime is a difficult task (e.g., aggregating logs)~\cite{li2015accommodate}. This involves a continual reauthentication process that requires cancelling all running tasks in order to start a new session. The new features of Hadoop 3.x need to include an adaptive mechanism for each expired token to continue to support the long-lived YARN applications service~\cite{long}.

\item Communication orchestration: There could be thousands of node-to-node communications for a job, resulting in the same magnitude traffic without a centralized (third-part) orchestration service that enforces authorization policies in a fine-grain model to guarantee data security across the Hadoop cluster.

 \item Data integrity: The communication channels between services still need to be secure; there is no end-to-end data encryption. This issue indicates that tokens need to be encrypted over the network (SSL/TLS may be utilized). Sniffing (sniffing attack) tokens are sufficient to incorporate an authorized entity, thus prompting a vulnerability to threaten the HDFS data.   
 
\item Node orchestration: Time needs to be roughly uniform across cluster nodes; else, the time-limited tokens will not work. Additionally, any compromised entity may manipulate the local time to infinitely extend the ticket lifetime.

\item Audit log management: It is difficult to analyse or test against the security configurations of each entity within a federation cluster. Different NNs separate log categories from regular processing logs and store them in separate locations with varying policies of persistence. 

\item Security auditing: It fails to cope with modern auditing demands of large-scale distributed clusters, as in a federation. Any secure IAM model should not only secure the access control but should be able to perform security auditing at the service level: for instance, incident reporting, behavioural analytics, and regular risk assessment of the cluster entities.

\item Access control scalability: Adding new security features, cluster entities, and updating policies, could require a complicated (and time-consuming) process that must be configured carefully. This model might be acceptable for a relatively small setup, but not for a federation infrastructure.
\end{enumerate}


\section{Federation access control reference model}
\label{RM}

\begin{sidewaysfigure}[!h]
  \centering
  \includegraphics[width=\linewidth]{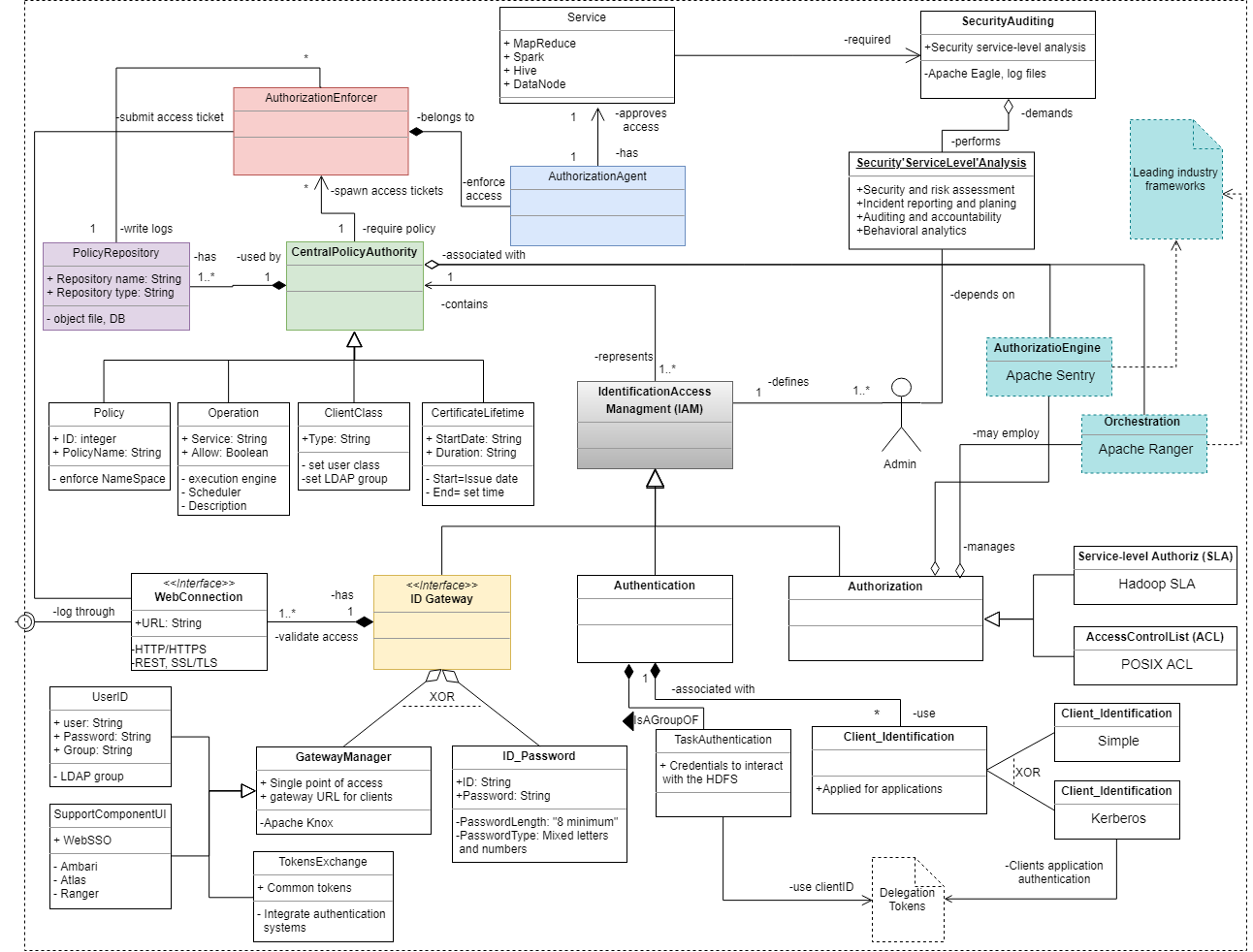}
  \caption{Federated Access Control Reference Model (FACRM) for Apache BD stack.}
  \label{fig:6}
\end{sidewaysfigure}

A service-oriented metamodel for access control support in BD federation platforms is presented in Figure~\ref{fig:6}. The Federated Access Control Reference Model (FACRM) is a set of architecture components that are associated with the security development of BD environments. FACRM also addresses many technical aspects that optimize context-specific design, which need to be clarified at different granularity levels. The FACRM is characterized by (i) ease of distributing secure access at scale, (ii) accelerated policy spawning and provisioning, (iii) providing ease-of-use horizontal scalability, and (v) flexibility for adapting to the service provider requirements and demands. Some highlighted elements will be discussed in Section~\ref{Summary}.

Next, we describe the FACRM components as listed below: 

{\bf IdentificationAccessManagement(IAM)}
Maintains the access control status of system entities within three layers of defense: ID gateway (to reach the cluster), authentication (identify verification; allowed to contact the underlying services), and authorization (which access rights and permissions are held).

a- GatewayManager: a registration authority for external login and all REST APIs. It represents the first IAM stage for verifying client service calls. It could employ either simple user name/password or a third-party framework, namely Apache Knox. By leveraging Knox as the cluster access point and URL gateway, it enforces access discussions behind the infrastructure firewall. Hence, failed requests neither reach the Hadoop daemon nor operate on its entities (e.g., NN and DN). In this scenario, Knox is composed of three main components to validate the access of several connections: 

 i- UserID: 
 identification obtained via profile (name/password) and Hadoop can set the group mapping service~\cite{Grouping}. However, resolving URL access via groups is managed via LDAP with simple authentication using JNDI-API.

\clearpage
\clearpage      
 
 ii- SupportComponentUI: Knox provides WebSSO capabilities to the Hadoop cluster. It is therefore essential to support different system component UIs and to link them to the end user. Examples include Apache Ambari UI for cluster development and orchestration and the Ranger admin console for setting/modifying policies.
 
    iii- TokensExchange: maintains a secure interaction among different frameworks and clients by providing a universal authentication platform that abstracts token exchange within federated Hadoop clients. 
    
    b- Authentication: the second stage of access control, which uses the client identification to authenticate the job service calls (after validating the client access to the cluster in the previous stage). The process takes place by specifying the core-site.xml in the Hadoop configuration files through enabling the hadoop.security.authentication feature. This configuration can be either simple or Kerberos authentication. In a multi NNs environment, like a federation environment, Kerberos manages the DNs authentications by defining the entities which are allowed to communicate with the HDFS manager.
    
    c- Authorization: the final stage of access control that sets each client permissions within the cluster entities. By default, this service is disabled; admins, however, can enable it using the Hadoop core capability in hadoop.security.authorization within the core-site.xml file. This includes the basic set of entity permissions using the SLA. An additional level of access control granularity can be acquired using HDFS POSIX ACL. For instance, ACL supports entity authorizations such as file-permission (read, write, execute). When the client creates a new file or sub-directory, it will automatically inherit the ACL permissions of the parent directory. However, clients are asked to gain separate authorization for each service. Correspondingly, client authorization (fine-grain access control) may be managed separately via different access control vendors, such as Apache Ranger and Sentry. Ranger provides dynamic data masking (in movement) for several frameworks of the Hadoop stack (e.g., HBase, Storm, Knox, Solr, Kafka, and YARN), while Sentry supports the Impala SQL query engine. The next section will demonstrate how ACL and Ranger/Sentry roles may be used interchangeably in a Hadoop cluster.  

{\bf WebConnection}: every client log has only one gateway, while the gateway manages several log connections.

{\bf CentralPolicyAuthority (CPA)}: a policy-based authority that keeps spawning access certification (tickets) for IAM and monitoring user access. It represents the central process of the IAM that releases access tickets and auditing functionalities. These tickets map each user/group with its granted permission and enforce service access discussions. It also sets the maximum lifetime and the start date of each policy, as well as the user class authorization intervals. The CPA validates the client login and creates an access certificate by checking both the Kerberos authentication list and Ranger authorization level access. Hadoop administrators may manage repository policies by setting up both Ranger and ACL policies (optional). In this case, Ranger will verify the fine-grain client access control, i.e., which HBase/Hive DB and table columns they have access to, Kafka queues, and HDFS level of access. Meanwhile, the ACL will verify the access control of remaining entities. However, Ranger policies will take priority over those of ACL. If a Ranger policy does not exist, then local ACL will take effect. Hadoop daemon authentications and internal communication (such as task status) will primarily rely on using the Kerberos principal and keytab file locations and are enforced using Hadoop core access control, i.e., ACL.
\\
  i- Policy: as a unique name for every policy, a policyID ensures that policies are not duplicated in the cluster.
  
   ii- Operation: defines the operation type (its authorization level value) and passes a Boolean that enables/disables the authorization level for every user/group for each service. Policies are enabled by default unless the admin restricts them.
   
   iii- ClientClass: assigns individual user permissions or sets group permissions for each policy, and defines each client class according to that policy.
   
v- CertificateLifetime: a validation interval of the certificate for the client session on the system.  

{\bf PolicyRepository}:
a file or DB that stores the certifications, as well as the auditing logs information. The repository functions may regularly include cache policies and track ticket updates (status saving). Numerous AuthorizationEnforcer daemons will write auditing logs to the PolicyRepository (as a unified auditing store). These policies can be classification-based, prohibition-based, time-based, and location-based.    

{\bf AuthorizationEnforcer}:
a process that enforces access decisions based on the CPA policies before allowing communication with its underlying services. It represents the NN daemon (see Figure~\ref{fig: 3}) in a native HDFS access control pattern. To generalize the IAM processes within different methodologies (i.e., only utilizing the Hadoop core capabilities or employing a third party), the presented metamodel separates this functionality. Every service call thus must pass through a predefined AuthorizationEnforcer before reaching the service.    

{\bf AuthorizationAgents}:
a distributed representative component that belongs to the AuthorizationEnforcer. Each service daemon (DN, HBase table) has an AuthorizationAgent that approves the access call. The AuthorizationAgent could be an ACL or a Ranger policy agent. 
 
 {\bf Service}:
the system processing framework, execution engines, and HDFS DNs. These services may include, but are not limited to: MapReduce for batch querying, Apache Spark for micro-batches, and Apache Storm for real-time processing. A comprehensive security ecosystem will require performing auditing measurements at the service level of BD applications and frameworks.  

{\bf SecurityAuditing}:
tracks the service calls and status of requests, in addition to monitoring client activities. This service may include identifying security and performance issues within the local log files or employing the Apache Eagle framework. These services require security auditing to improve the overall security measurements.  

{\bf SecurityAnalysis}:
carrying out a security assessment based on client behavioural analysis, incident reports, and audit pieces of information. 

{\bf Admin}:
the one who sets the policy for other users of the system. The admin depends on the SecurityAnalysis and client demands to define the IAM policies. Admin is also responsible for securely installing/configuring the cluster entities, maintaining a secure operational environment, performing security patching policy, and scaling computing resources based on peak utilization.


\section{Implementation and Validation}
\label{Implementation}

In general, BDF provides a systematic way to dynamically provision the user demands with elastic resources and services over shared data blocks. Implementing federated authorization and delegation to enhance BD client access decisions is vital for the success of adopting these new features of Hadoop 3.x. In this paper, we argue that the FACRM is essential for the concept of unifying the IAM in a BDF infrastructure and the data that they stock. We demonstrate how the proposed components of BDF deployments can be brought to a secure access broker architectural pattern. This BDF access broker abstracts the identification, authentication, and authorization discussion within the large-scale data analytics of the Hadoop 3.x platform. Furthermore, we explain the required steps to utilize the FACRM as a multi-tier Hadoop architecture within any deployment architecture.

\subsection{Federation Access Broker}
\label{Broker}

\begin{figure}[!t]
  \centering
  \includegraphics[width=.95\linewidth]{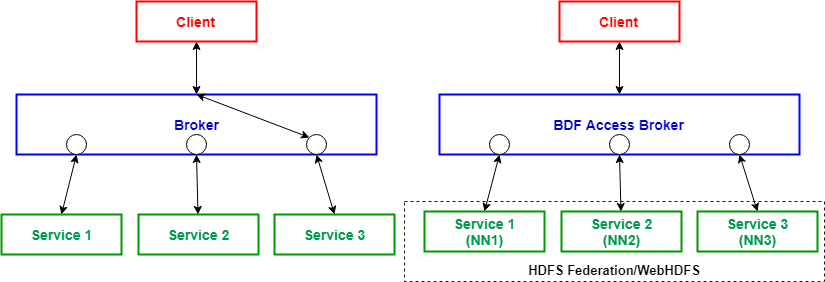}
  \caption{A high-level abstraction of general access broker pattern vs. the proposed BDF access broker}
  \label{fig:broker1}
\end{figure}

Conventionally, system architectures employ an intermediate layer, i.e., intermediate brokers that orchestrate the underlying service’s communications with external entities. The most common implementation of such architectures relies on forwarding all of the client’s traffic into a proxy server or gateway for authentication. Such applications are called access brokers, which provide perimeter security by adding an authorization layer based on policies. An access security broker is thus an on-premise or cloud-based security policy enforcement point. It is placed between the external client and the underlying services provided by the system.

In general, the broker architectural pattern can be used to structure distributed systems with decoupled components that interact through remote service invocations. In this study, we present and demonstrate a pattern for this type of system, and we highlight its importance on the federation scale. Figure~\ref{fig:broker1} shows a conventional access broker (on the left) and an abstracted architecture of the proposed BDF access broker. The proposed BDF access broker aims to abstract the management of the remote access calls to Hadoop federation services and data without user intervention. Thus, it is responsible for coordinating security policies (authentication and authorization) as BDF assets are accessed (data, services, and infrastructure). It also enables centralized control through the utilization of a unified pattern, which facilitates auditing and analysis.

\begin{figure}[!t]
  \centering
  \includegraphics[width=.75\linewidth]{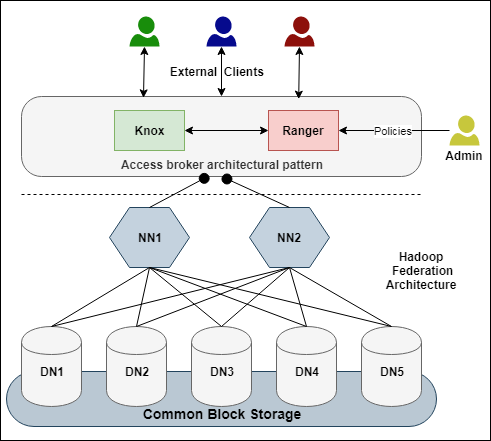}
  \caption{BDF access broker architecture pattern}
  \label{fig:broker2}
\end{figure}

The proposed architecture broker pattern in Figure~\ref{fig:broker2} is composed of (i) authentication (credentials and passwords) using Apache Knox, (ii) authorization policy enforcement using Apache Ranger, and (iii) security analysis/auditing capabilities using a centralized access audit log repository and Apache Hadoop 3.x. In particular:

\begin{enumerate}
\item  Knox gateway access: employed for external client identification and cluster access.

\item Ranger authorization assignment: the clientID is passed to Ranger to validate the access call to contact NNs, and a secure session is established. If successful, Ranger responds with a certificate for authorization level access and assigns the permissions of each NN (defined by the admin) to reach the stored data in the DNs which belong to that specific NN. 

\item Security log: access audit logs are created with each attempt to reach the data and are stored in unified HDFS storage (see Section~\ref{Audit}).
\end{enumerate}

\begin{table}[t!]
\centering
\caption{Federation Supported Hadoop Access Control Model}
\label{tab c1}
\resizebox*{!}{\totalheight}{%
\begin{tabular}{@{}llllll@{}}
\toprule
\multicolumn{6}{l}{\textbf{Basic Sets and Functions}}\\
\multicolumn{6}{l}{-- U, G and S (finite set of users, groups and subjects, respectively)}   
\\
\multicolumn{6}{l}{-- HS, \ophs (finite set of Hadoop services and operations, respectively)} 
\\
\multicolumn{6}{l}{-- \dirug~ : $ \mathrm{U \rightarrow 2^{G}} $, mapping each user to a set of groups,  equivalently UGA $ \subseteq U \times G $} 
\\
\multicolumn{6}{l}{-- $\mathrm{\hsprms = 2^{HS \times \ophs}}$, set of Hadoop service permissions}                                                                                                          \\
\\

\multicolumn{6}{l}{ \textbf{Permission Assignments}}                                                                   \\
\multicolumn{6}{l}{\begin{tabular}[c]{@{}l@{}}-- $ \mathrm{\pahs \subseteq (U \cup G) \times }$\hsprms,  mapping entities to Hadoop service permissions. \\ Alternatively,\\ \hsp~: (x) $\rightarrow  {2^{\hsprms}}$, defined as \hsp(x) = \{p $|$ (x,p) $\in$ \pahs, x $\in$ (U $\cup$ G)\} \end{tabular}} \\
\\
\multicolumn{6}{l}{ \textbf{Effective User Permissions}}                                                                                          \\
\multicolumn{6}{l}{\begin{tabular}[c]{@{}l@{}}$ \;\;\;\; \bullet \;\; \effhp : \U \rightarrow {2^{\hsprms}}$, defined as \\\;\;\;\;\;\; $ \effhp(\mathrm{u}) = \hsp(\mathrm{u}) $ $\cup $  $ \mathrm{\bigcup\limits_{ g \; \in\; \{\dirug(u) \}} \hsp(g)}$\end{tabular}} \\[\defaultaddspace]

\multicolumn{6}{l}{\textbf{User Subject}}\\
\multicolumn{6}{l}{\begin{tabular}[c]{@{}l@{}}$\;\;\;\; \bullet $\; \userSub~: \s $\rightarrow \; $U, mapping each subject to its creator user, where the subject \\\;\;\;\;\;\;\;\;acquires some or all of the permissions of the creator user.\end{tabular}}                                                                                                          \\
\midrule
\multicolumn{6}{l}{\textbf{Hadoop Service Access Operation Decision}}\\

\multicolumn{6}{l}{\begin{tabular}[c]{@{}l@{}}A subject $\mathrm{s}$ $\in$ S is allowed to perform an operation op $\in$ \ophs on a service hs $\in$ HS \\ if the effective permissions of \userSub($\mathrm{s}$) include permission assignments for hs $\in$ HS.\\ Formally,
(hs,op) $\in$ \effhp($\mathrm{\userSub(s))}$

\end{tabular}}                                                                                      \\

\bottomrule
\end{tabular}
}
\end{table} 


\subsection{Formal Access Control Model Components}
We have formally defined the federation supported access control model in Table \ref{tab c1}. This model has been adapted from the object tagged RBAC model \cite{gupta2017object} for Hadoop, and introduces required components to demonstrate the federation.

The basic components for the model include a finite set of Users (U), Groups (U), and subjects (S), which are created by the user and run on its behalf. A user can belong to groups, and permission can be assigned to groups which trickle to the member users. 
Hadoop Services (HS) are the required daemon background processes, including NameNode, DataNode, YARN ResourceManager, etc., to which the user must be allowed access. In the case of federated Hadoop 3.0, because we have multiple NNs in the system, the user must have access to only specific ones. \ophs are the actions of Hadoop services. ACLs have been used in native Hadoop service authorization capabilities to restrict access to users.

As shown in the table, the Hadoop service permissions (\hsprms) are the power set of the cross products of HS and \ophs. A many to many relation \pahs specifies the assignment of Hadoop service permissions to the users or groups. In this way, a user can be assigned permission to access Namenodes in the system, either directly or through group membership (using the sufficient permissions \effhp). For example, a user u1 may only be allowed to access Namenode namenode1, and then the permission assignment must be only for that tuple (u1,namenode1) in the \pahs relation. 

A user creates a subject stated by the \userSub function. The subject obtains either some or all of the permissions of its creator user. A subject is allowed to access a Hadoop service if its creator user has the permissions assigned to it by the security administrator.


\begin{table}
\begin{adjustwidth}{-1in}{-1in}  
\begin{center} 
\caption{BDF access Broker proof of concept validating environment.}\label{tab:experiment}
\centering 
\scalebox{.75}{
\begin{tabular}{|c|c|c|} \hline
\rowcolor{lightgray}  \textbf{Node Number} &  \textbf{Hadoop Daemon} &  \textbf{ Hardware \& \newline Configuration} \\
\hline
2 & NameNode &
      VCPUs:    8,
        RAM:    16GB,
        HDD:    160GB \\ \hline
3 & DataNode & 
      VCPUs:    4
        RAM:    8GB
        HDD:    80GB  \\ \hline
\multicolumn{3}{|c|}{Common criteria: 1 Gbps network connection, Ubuntu 16.04 LTS OS.}  \\ \hline
\end{tabular}}
\end{center}
\end{adjustwidth}
\end{table}


\subsection{Experiment and results}
\label{Results}

\begin{figure*}[h!]
    \centering
    \subfloat[WebHDFS Read ]{{\includegraphics[width=5.5cm]{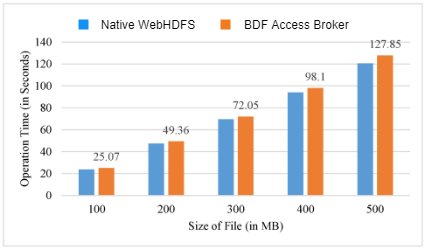} }}%
    \qquad
    \subfloat[WebHDFS Write]{{\includegraphics[width=5.5cm]{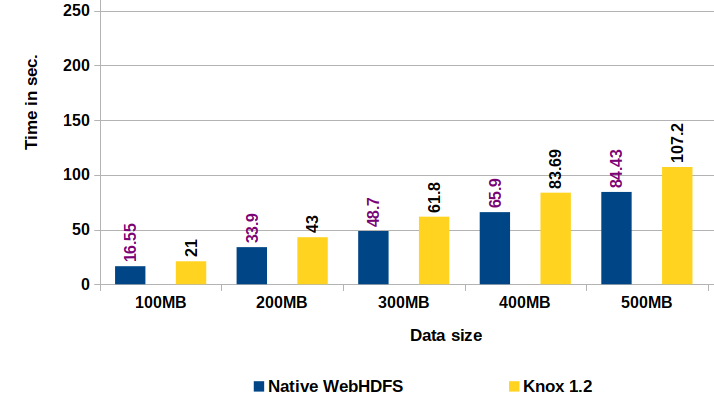} }}%
    \caption{Performance analysis of WebHDFS operation within different file sizes}%
    \label{fig:example1}
    \subfloat[WebHDFS Read)]{{\includegraphics[width=5.5cm]{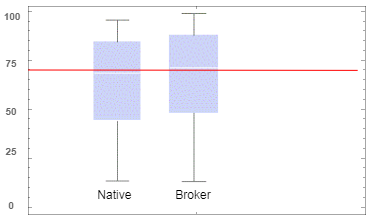} }}%
    \qquad
    \subfloat[WebHDFS Write]{{\includegraphics[width=5.5cm]{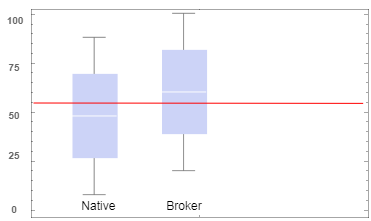} }}%
    \caption{A box and whisker plot displays a performance comparison of BDF broker against native (non-secure) Hadoop implementation}%
    \label{fig:example2}
\end{figure*}

Aiming to validate the BDF access broker, we implement an OpenStack~\cite{sefraoui2012openstack} based Hadoop cluster that is configured as one rack of five nodes using two different node types, as detailed in Table~\ref{tab:experiment}. The bar chart in Figure \ref{fig:example1} illustrates a performance comparison of the WebHDFS read operation in two use cases: the first without utilizing any security measurements (native WebHDFS access), and the second by employing the proposed BDF access broker architecture pattern. The prime motivation for this performance evaluation is to gain an understanding of how the performance will be impacted with respect to Knox and Ranger. Overall, it can be observed that the performance effect, i.e., the time needed to process different file sizes, seems to be minimal. A more specific study on performance degradation is eventually needed, but for this proof-of-concept, it is correct to state that the performance has not been significantly impacted.

To define the relationship between the basic WebHDFS configuration and secure WebHDFS configuration with the proposed broker, we compare the speeds of the two methods. We fitted two linear regression models (see Appendix A) that can be found in Table ~\ref{tab:Write} and ~\ref{tab:Read}. In the first case, the slope is approximately 0.24, and 0.25 in the second. These results show that the time increases by only 1\% for each MB using the secure approach. Figure \ref{fig:example1} also shows the time needed to perform WebHDFS calls (read/write) over different data points (data chunks between 100 MB and 500 MB). 

\begin{table*}[h!]
\caption{Access Control impact on Read performance with Sample Standard Deviation (SSD) and Transfer Speed (TS)}
 \centering
 \scalebox{0.75}{
\begin{tabular}{l|l|l|l|l|l|l|l|l|l|}
\cline{2-10}
                                     & 100MB & 200MB & 300MB & 400MB & 500MB    & TS & Mean & Median & SSD* \\ \hline
\multicolumn{1}{|l|}{Native WebHDFS} & 23.64 & 47.45 & 69.63 & 94.23 & 120.62 & 25 MB/s & 71.11 & 69.63 & 38.08       \\ \hline
\multicolumn{1}{|l|}{BDF Access Broker}       & 25.07 & 49.36 & 72.05 & 98.1 & 127.85 & 26 MB/s & 74.49 & 72.05 & 40.26          \\ \hline
\end{tabular}}
 \label{tab:Write}
\end{table*}

\begin{table*}[h!]
\caption{Access Control impact on Write performance with Sample Standard Deviation (SSD) and Transfer Speed (TS)}
 \centering
 \scalebox{0.75}{
\begin{tabular}{l|l|l|l|l|l|l|l|l|l|}
\cline{2-10}
                                     & 100MB & 200MB & 300MB & 400MB & 500MB & TS & Mean & Median & SSD* \\ \hline
\multicolumn{1}{|l|}{Native WebHDFS} & 16.55 & 33.9  & 48.7  & 65.9  & 84.43 & 42 MB/s & 49.9 & 48.7 & 26.54       \\ \hline
\multicolumn{1}{|l|}{BDF Access Broker}       & 21    & 43    & 61.8  & 83.69 & 107.2 & - & 63.34 & 61.8 & 33.71          \\ \hline
\end{tabular}}
 \label{tab:Read}
\end{table*}


\begin{figure*}[h!]
    \centering
    \subfloat[WebHDFS Read ]{{\includegraphics[width=5cm]{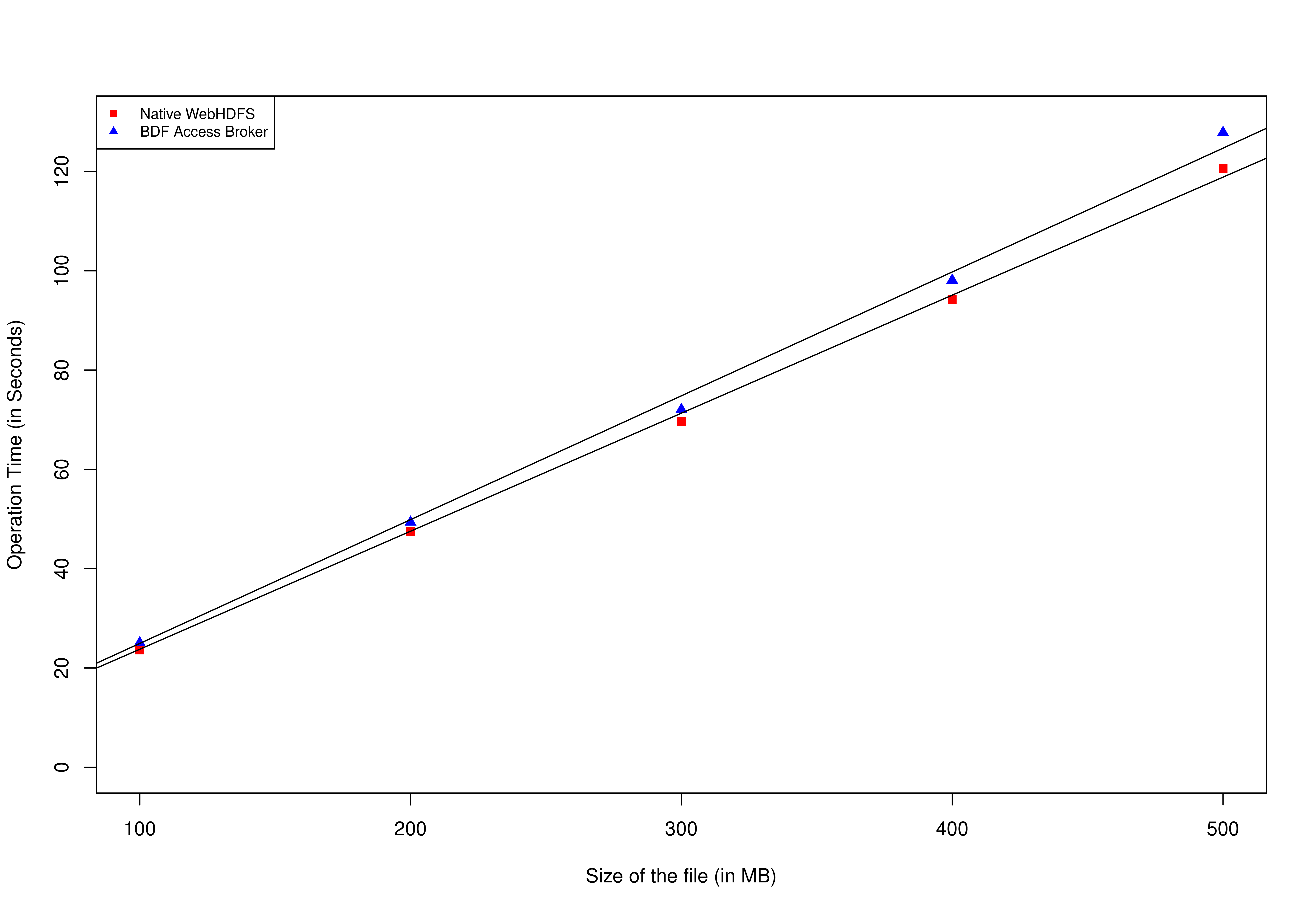} }}%
    \qquad
    \subfloat[WebHDFS Write]{{\includegraphics[width=5cm]{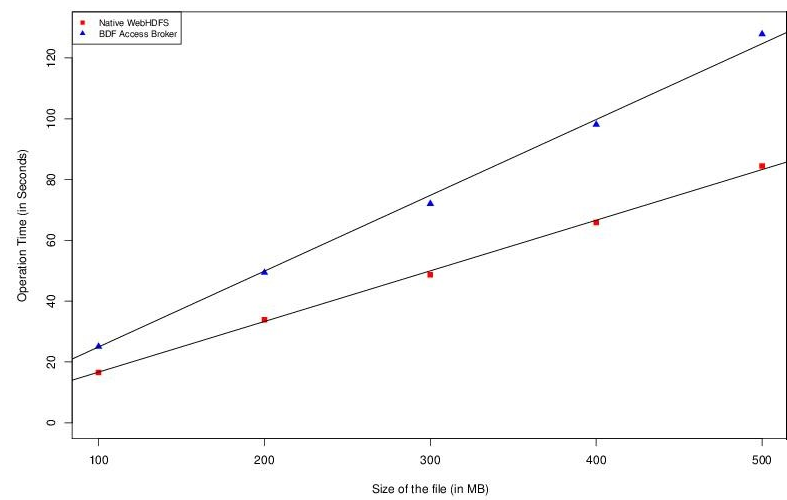} }}%
    \caption{Graph of the relationship between secure and non-secure WebHDFS rates}%
    \label{fig:example3}
\end{figure*}
A box and whisker chart in Figure ~\ref{fig:example2} shows the distribution of data into quartiles, highlighting the mean and outliers. Our experiment shows the box and whisker plot medians, interquartile ranges, and ranges of the secure and non-secure approaches. This experiment aimed to identify covariates that could influence the length of time taken to perform a read/write call by using the box and whisker plots to allow visual comparison of the median and spread. Areas of potential interest are highlighted for future prospective work. The whiskers extend from the box to the highest and lowest values, and our experiment shows that no outliers exist. Also, Figure~\ref{fig:example3}, shows that the relationship between the two variables is positive linear.

\section{Access Audit Log Management and Analysis}
\label{Audit}

NN, as a data lake gateway, writes access audit log information to local files after each successful/failed access call. These access log files generally contain information regarding the user and the data the user accessed. Typically, Hadoop creates logs after each user read/write call using Apache Log4j files~\cite{log4}, i.e., a Java library. This detailed log information is managed locally by each Hadoop component and provides a useful accounting process (e.g., timestamp and IP address) that logs all operations. The log files may serve multiple purposes, such as debugging operational issues and regulatory compliance. They also provide security capabilities to execute forensics from an access audit perspective (e.g., user logs over HTTP). In particular, access audit logs are valuable pieces of information for conducting security auditing and analysis of the Hadoop cluster with high operational resolution.

However, analysing the collected YARN and HDFS logs generated by applications of many distributed components (locally at each daemon) is a difficult and insufficient task. Each Hadoop daemon creates its own output log messages and manages them in a local directory (on the node on which it is running). Hadoop Archives may be employed to reduce the number of small files and thus the stress on the NN~\cite{archive}. However, it is still difficult to perform security analysis or digital forensics over a large-scale Hadoop federation model that could include dozens of NNs and hundreds of DNs with broad user participation. Another way to reduce the complexity of the distributed log file analysis is by utilizing log aggregation: for instance, YARN log aggregation, a management process that fuses different log files from various sources in a centralized repository to facilitate the analysis of the collected logs. This feature is disabled by default; by enabling it, the log files of different daemons will be allocated to a centralized directory (e.g., HDFS). However, even using this feature, the large amount of aggregated data in Hadoop federation architecture will lack governance requirements, like file tagging, labeling, and classification. It will be stressful for investigators to examine such pieces of information, and it can easily frustrate them.

\begin{figure*}[t]
  \centering
  \includegraphics[width=.7\linewidth]{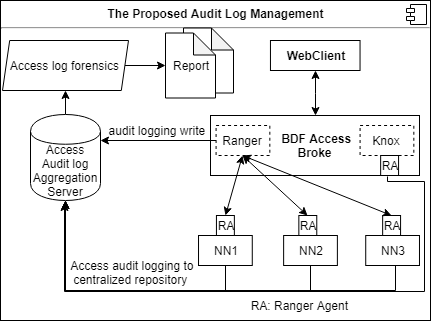}
  \caption{A conceptual BDF access broker log management model}
  \label{fig:9}
\end{figure*}

The proposed access broker not only provides fine-grained access control but also a centralized auditing approach (see Figure~\ref{fig:9}). By employing the Ranger Audit Server, the BDF access broker will aggregate all access logs into a centralized repository (RDBMS, HDFS, or Log4j). The log aggregation service will regularly check the running tasks and start to aggregate and transfer the logs to a centralized repository (typically HDFS). Subsequently, access audit logs can be allocated by other digital forensics tools for any visualization or search tool for further investigation. Moreover, server audit logs provide valuable information to identify and repair system vulnerabilities.

Consider the following scenario that illustrates how to investigate a security breach using the access audit logs by storing and analysing the access log data.

A Hadoop system administrator notices a tenfold spike in the number of support tickets. Because the system utilization is high, due to the unusual ticket requests, clients start facing problems logging into the VPN. The admin must analyse the log data, since he suspects that he may be experiencing a distributed denial of service (DDoS) attack. The questions that he must answer are: “how many connection requests have been delivered in the past few hours?”, “where are the requests coming from?”, and “what types of connections are they (authorized or not)?” The analysis includes investigating how and when the network traffic changed. Doing so requires engaging in three steps to transform the logs into useful figures, namely, loading, refining, and visualizing the data, in the search for useful information, as follows:

\begin{itemize}
    \item Loading log data: Utilizes Apache Flume~\cite{hoffman2015apache} to stream large amounts of log data from the VPN server into the HDFS. The log records contain a timestamp, IP address, country, and indicator about whether the connection was successful or not.  

\item Refining log data: A storage management layer must be employed to enable a relational view of data without affecting the data format. This layer aids in browsing the log data (making sure it has been loaded correctly). Apache HCatalog, i.e., Hadoop catalogue, may be utilized to provide a metadata layer and rest interface for Hadoop data. Next, the admin needs to write them onto any data analytics framework for review (visualization tool). An Apache Pig script may be used to push the latest log data to the visualization tool. 

\item Visualizing log data: The admin interfaces the visualization tools, like ElasticSearch (a real-time search engine and analytics tool), and Microsoft Excel (using open database connectivity). ElasticSearch provides high-level visualization to identify the moment when the network traffic increased, where it came from, and whether or not it was authorized in order to confirm his hypothesis about the DDoS attack. More detailed maps may be created using the Excel Power View functionality by linking the Apache Hive (the base of HCatalog) and Excel to determine whether there is cause for concern.

\end{itemize}

The forensic report and the process output data are used to update the firewall to deny requests from the unauthorized IP addresses. To automate this update, the admin can use Apache Ozzie to schedule a job to automatically update his firewall every hour. This report can also be used for other forensics to respond to different security threats, prepare for compliance audits, and prepare behavioural analytics (for security patching).


\section{Summary and Future directions}
\label{Summary}

Next-generation access control systems aim to become more scalable with even finer-grained authentication and authorization management~\cite{awaysheh2019poster}. However, given the emerging trends of the distributed, heterogeneous, and federated infrastructure associated with Hadoop 3.x architectures, the need for a lightweight security framework is inevitable. In this paper, we present a BD Federation-oriented reference model for the secure development of access control solutions within Hadoop clusters. The proposed model is a generic high-level conceptual metamodel (called the FACRM) that aims to formalize Hadoop 3.x core access control capabilities and highlight some of Hadoop’s top-level security projects. It also aims to successfully implement a unified and secure BD solution over on-premise and cloud-based deployment architectures in favour of BD applications and frameworks. The FACRM supports myriad access control options across a variety of policies, users, and frameworks in pursuit of modern BD infrastructure. Furthermore, this study presents an in-depth investigation of the Hadoop IAM ecosystem and draft current access control limitations within a federation deployment architecture. We validate the utilization of our proposed RM by implementing a proof-of-concept architecture for the security and privacy of BDF deployment. Our test conclusively confirms the effectiveness of the proposed access pattern performance and security.  

The general aim of this study is to develop an RM using open-source frameworks (e.g., Knox, Ranger) to promote the adoption of reliable access control mechanisms within the BD deployment architectures. It also aims to facilitate the security design of all clients, resources, applications, and networks in Hadoop-based BD processing environments. In this context, the proposed federation BD access control meta-model is a service-oriented architecture (SOA) that conforms to the best practices of independent vendors, products, and technologies. This SOA makes FACRM frameworks (a framework design based on FACRM) into SOA frameworks. However, FACRM frameworks are not necessary for BD frameworks only; it could be any framework or application that runs over the HDFS system, which is a large-scale distributed environment (platform). This environment is, thus, characterized by horizontal scalability, rapid provisioning of access control, ease of access, and security. Additionally, Our BDF access broker provides an attractive solution to the audit log problem that effectively addresses compliance requirements. To map the current Hadoop limitations (presented in Section \ref{limitation}) to the solution domain of this study, Table \ref{tab:Model} represents these limitations as threats and links them with solutions.

\begin{table}[ht!]
\begin{adjustwidth}{-1in}{-1in}  
\begin{center} 
\caption{Linking current Hadoop federation limitations to the proposed solutions}\label{tab:Model}
\centering 
\label{tab:summary}
\scalebox{0.75}{
\begin{tabular}{|p{3cm}|p{3.5cm}|p{10cm}|}\hline
\rowcolor{lightgray}  \textbf{Current \newline limitations} &  \textbf{Proposed solution} &  \textbf{Description}\\
\hline

Access control \newline complexity & FACRM propose that policies are written once and applied many times. & In Figure 6, the CentralPolicyAuthority (in green) allows the admin to set up a policy and distribute it to all nodes (using the Ranger agents), which are represented by the AuthorizationAgent (in light blue) attached to all of the services.  \\ \hline

Long-running services & FACRM propose flexible time-based policies. & The CertificateLifetime in the proposed reference model corresponds to the long-running services. These certificates are associated with the CentralPolicyAuthority and saved at the PolicyRepository to be updated automatically upon demand. \\ \hline

Communication \newline orchestration & FACRM propose a third-party centralized access control. & Security verification for all of the entities (users, nodes, and daemons) inside the Hadoop cluster and external clients. In Figures 7 and 8, it can be observed that the broker transfers the access discussion responsibility from the NN to the broker. \\ \hline

Data integrity & access control at the file level and service level (access call). & A federation configuration requires authorization verification of the entire file path to create secure multi-tier federation architecture. Additionally, multi-layer control (authentication and authorization) occur at the beginning of operation (reading and writing to HDFS). \\ \hline

Node \newline orchestration & FACRM provide access control using any approach. & FACRM and BDF broker provide the base to support the multipart access control framework and technologies, e.g., RBAC, ABAC, etc. \\ \hline

Audit log \newline management & utilization log aggregation using centralized auditing approach. & Figure 12 illustrates how to employ the proposed broker with Ranger Audit Server for accessing broker log management.  \\ \hline

Security \newline auditing & proposal of an architecture pattern for BD governance. & In Section 7, we proposed a scenario that transforms the logs into useful figures by loading, refining, and visualizing the data into incident and forensic reports using open-source frameworks. This section shows how to investigate a security breach using Apache Flume, Apache HCatalog, and ElasticSearch, among other tools.\\ \hline

Access control scalability & proposal of a novel BDF access broker. & The main components of our proposed broker provide high scalability. In particular, Ranger provides centralized Hadoop security administration and management, while Knox streamlines security for services and users who access the cluster data and execute jobs.\\
\hline
\end{tabular}}
\end{center}
\end{adjustwidth}
\end{table}
\clearpage

Overall, the primary benefits of the BDF broker architectural pattern can be summarized in four different categories:

\textbf{Enhanced security:} A method to ensure that data is stored securely in BDF and BDaaS clouds using a robust access control mechanism. This involves using (i) Knox to expose REST and HTTP services without revealing the details of the underlying Hadoop cluster, and (ii) Ranger for fine-grain authorization over federation NNs.

\textbf{Centralized control:} Using a single gateway with a distributed agent to prevent unauthorized access to the modern data lake. This also allows IT departments to set and enforce security policies regarding data usage, access calls, and resource utilization over secure channels.

\textbf{Facilitated auditing and analysis:} A centralized repository for all access logs with a sophisticated governance and accountability approach (reporting and regulatory compliance). This approach enables threat prevention methods, e.g., behavioural analytics and threat intelligence.

\textbf{Consolidated governance pattern:} A way to ensure and prove compliance with granular visibility and control to meet regulatory requirements, and to guide the efforts of admins to ensure that the system complies with all relevant regulations and standards (such as data residency) when using Hadoop 3.x.

Blockchain technology can serve as a revolutionary solution, addressing current BD privacy concerns such as wire (end-to-end) encryption and decentralization data-driven cryptography.
As an enabling technology, blockchain provides a sequence of block cryptography that stores all transactions, which has been established as a solution for the large-scale distributed agents of Bitcoin system security~\cite{khan2018iot}. The core functionality of blockchain technology relies on providing robust cryptographic (each agent is assigned a private key, whereas a public key is shared with all other agents) proof for data authentication and integrity by providing a list of all transactions and a hash to the previous block. This list is verified by majority agreement of nodes (P2P network where the sender broadcasts it to all of the other nodes) that are actively involved in verifying and validating transactions~\cite{eyal2016bitcoin}.

The current Hadoop setup does not enforce access control by the DN to access its data blocks. Another issue remains in that both NN and DN generate a block access token using the same key. If an attacker leaked the key, he could generate a token and access any data blocks in the HDFS. Also, the leaked DT can be used to steal a large amount of data from HDFS, since it has the privilege to behave as the Hadoop client and to access all content which he is allowed to access. Any future BD Hadoop solution should consider the previous limitations on a federation scale. To this end, FACRM addresses all concerns mentioned above by defining a reference model for Hadoop federation; more particularly, it presents a meta-model that includes the main access control vocabulary and design elements, the set of configuration rules, and the semantic interpretation. Leveraging recently published management mechanisms of policy enforcement, e.g. ~\cite{ahmad2018lazy}, to combine cloud and edge federations with role-based and attribute-based access control~\cite{gupta2017multi, gupta2018attribute} could be considered in future research. Examples of different research directions include:

\begin{itemize}
    \item \textbf {Hybrid access control solutions:} The development in usage control for data privacy and security leads to integrating conventional access control approaches, e.g., ABAC, with other context-based encryption technologies. This integration could include functional encryption or other generalizations of identity-based encryption and attribute-based encryption for protecting big data in the presence of mutable attributes, which may lead to general advancement in other fields, such as IoT \cite{aladwan2019common}.
    
    \item  \textbf {Fine-grained Access Control:}  The need to develop active and fine-grained approaches (policies and standards) for BDF is inevitable. The challenge here is to extend the previous approaches with new policy editing and presentation tools (classification-based, prohibition-based, time-based, and location-based policies) for flexible and extensible data access. In particular, this includes policies that extend ABAC with stateful access sessions and mutable attributes (i.e., characteristics that dynamically change during the session). This extension is invaluable to advance the authorization mechanisms of the multi-tenant data lake architecture, as well as the latest BDF model. Tackling this issue can be achieved by deploying policy templates that extend the primary ACL (or any access decision enforcer) as a specialization of the XACML V3.0 standard.  

    \item \textbf {Dynamic Access Control:} More than ever, scalability and dynamicity of the BD deployment architecture in terms of user and node participation are vital for the secure delivery of security provisioning in a federation environment. There have been different adapted security mechanisms and techniques proposed to cope with this concern. However, more research activities are still needed to establish scalable security models and paradigms that must be driven by BD specifications and requirements. Also, new security abstractions for BDF are still needed to simplify the task of identifying the unimproved gaps. For instance, leveraging new mechanisms of policy enforcement~\cite{ahmad2018lazy} to combine BDF with ABAC ~\cite{gupta2018attribute} could be pursued in future research.
    
    \item\textbf {Stateful Access Sessions:}  This requires restricting the BDF access to only those authorized by the succinct gateway. However, the current state-of-the-art BD-based encryption technologies involve transparent data encryption. Federation environments track the participation states and characteristics of the clients and networks traversing them with mutable attributes (i.e., the values of the attributes change over time during an access session). This operation requires formulating policy templates on top of standards such as XACML V3.0 for dynamic authorization with stateful access sessions. This takes place at the service level (e.g., data access calls) and does not protect data-in-transit or data-in-process (not even the metadata). Attribute-based encryption in conjunction with the usage control approach and ABAC can be labeled as a future direction for this issue. 
    
\end{itemize}

Providing new solutions for all of these research directions will promote innovative BDF solutions with advanced security features. It is also expected that adoption of the FACRM use cases will be implemented within our BD opportunistic and elastic resource allocation (OPERA) platform. OPERA architecture (prototype proposed in~\cite{awaysheh2017eme}) combines the computing power of high-throughput resources available (non-dedicated) to the Hadoop 3 dedicated cluster using Docker containers as worker nodes. Those non-dedicated containers tend to be more vulnerable, which requires a robust IAM approach to minimize security threats.

\section*{Acknowledgement}

This work has received financial support from the Conseller{\'{\i}}a de Cultura, Educaci{\'o}n e Ordenaci{\'o}n Universitaria (accreditation 2016-2019, ED431G/08 and reference competitive group 2019-2021, ED431C 2018/19) and the European Regional Development Fund (ERDF). Additional support was received from the Ministerio de Econom{\'{\i}}a, Industria y Competitividad within project TIN2016-76373-P (AEI/FEDER, EU), the Xunta de Galicia GRC R2014/008, networks R2016/045, R2016/037 and CAPAP-H, and the \textit{Conseller{\'\i}a de Cultura, Educaci{\'{o}}n e Ordenaci{\'{o}}n, Universitaria de la Xunta de Galicia} (2016-2019, ED431G/08) and ERDF.

\bibliography{mybibfile}

\end{document}